\documentclass[12pt]{article}
\usepackage{ascmac}
\usepackage{latexsym}
\usepackage[dvipdfmx]{graphicx}
\usepackage{amsmath}
\usepackage[top=3.5cm,bottom=3.5cm,left=2cm,right=2cm]{geometry}
\usepackage{pifont}          
\usepackage{amsmath,amssymb}
\usepackage{color}
\usepackage{geometry}
\usepackage{caption}
\usepackage{ulem}
\newcommand{\h}{\hspace}
\newcommand{\be}{\begin{equation}}
\newcommand{\e}{\end{equation}}
\newcommand{\aln}[1]{\begin{align}#1\end{align}}

\begin{document}

\title{
\vbox{
\baselineskip 14pt
\hfill \hbox{\normalsize KUNS-XXXX
}} \vskip 1cm
\bf \Large  
Classical conformality in the Standard Model\\
 from Coleman's theory  
\vskip 0.5cm
}
\author{
Kiyoharu~Kawana\thanks{E-mail: \tt kiyokawa@gauge.scphys.kyoto-u.ac.jp}
\bigskip\\
\it \normalsize
 Department of Physics, Kyoto University, Kyoto 606-8502, Japan\\
\smallskip
}
\date{\today}

\maketitle

\abstract{\normalsize
The classical conformality is one of the possible candidates for explaining the gauge hierarchy of the Standard Model. We show that it is naturally obtained from the Coleman's theory on baby universe.
}
\newpage

Although the Standard Model (SM) is completed by the discovery of the Higgs boson, there are many open questions in it such as the Higgs quadratic divergence, the origin of the electroweak symmetry breaking (EWSB), the cosmological constant problem, and so on. The first two are important in that sense that they are inherent in local field theory even without considering cosmology. Therefore, it is quite important to seek for new theory or mechanism that naturally answer these questions.

The classical conformality (CC) \cite{CCS} 
recently attracts much attention as one of the candidates. 
The basic idea is as follows: The bare Higgs potential is given by 
\aln{
V_B^{}(h_B^{})&=\frac{A_B}{2} h_B^2+\frac{\lambda_B^{}}{4}h_B^4
\\
&=\frac{A(\mu)}{2} h^2+\frac{\lambda(\mu)}{4}h^4+\frac{\delta A(\mu)}{2} h^2+\frac{\delta \lambda(\mu)}{4}h^4
,\label{eq: bare potential}
}
where $\mu$ is the renormalization scale, $A_B^{}$ $(A(\mu))$ and $\lambda_B^{}$ $(\lambda(\mu))$ are the bare (renormalized) couplings, and $\delta A(\mu)$ and $\delta \lambda(\mu)$ are the counter terms. By using the wave function renormalization $Z_h^{}$, they obey 
\be Z_h^{}A_B^{}=A(\mu)+\delta A(\mu)\ ,\ Z_h^2\lambda_B^{}=\lambda(\mu)+\delta\lambda(\mu).
\e
As is well known, $\delta A(\mu)$ contains the quadratic divergence, so we must drastically tune the bare mass term $A_B^{}$ to obtain $-A(\mu)\sim{\cal{O}}(100\text{GeV})$. However, such a fine-tuning is automatically taken place if we demand that $A(\mu)$ becomes zero at some scale $\Lambda$:
\be 0=A(\Lambda)(=Z_h^{}A_B-\delta A(\Lambda)).\label{eq:cc}
\e
Once this is satisfied, the quadratic divergence never appears in the physical quantities 
. 
This is the basic idea of the CC. The purpose of this paper is to show that Eq.(\ref{eq:cc}) is naturally concluded from the Coleman's theory. Note that, in the SM, because the renormalization group equation of $A(\mu)$ is proportional to itself, the Higgs mass term never appears if the CC is realized at some scale. Therefore, other mechanisms such as the CW mechanism are needed to realize the EWSB, and they are well studied in the context of the models beyond the SM \cite{CCS}. In such extensions, a few scalars generally exist, and we do not yet understand whether our (following) argument is also applicable to such general cases. In this paper, we just concentrate on the CC in the SM.
%

\begin{figure}
\begin{center}
\includegraphics[width=9.5cm]{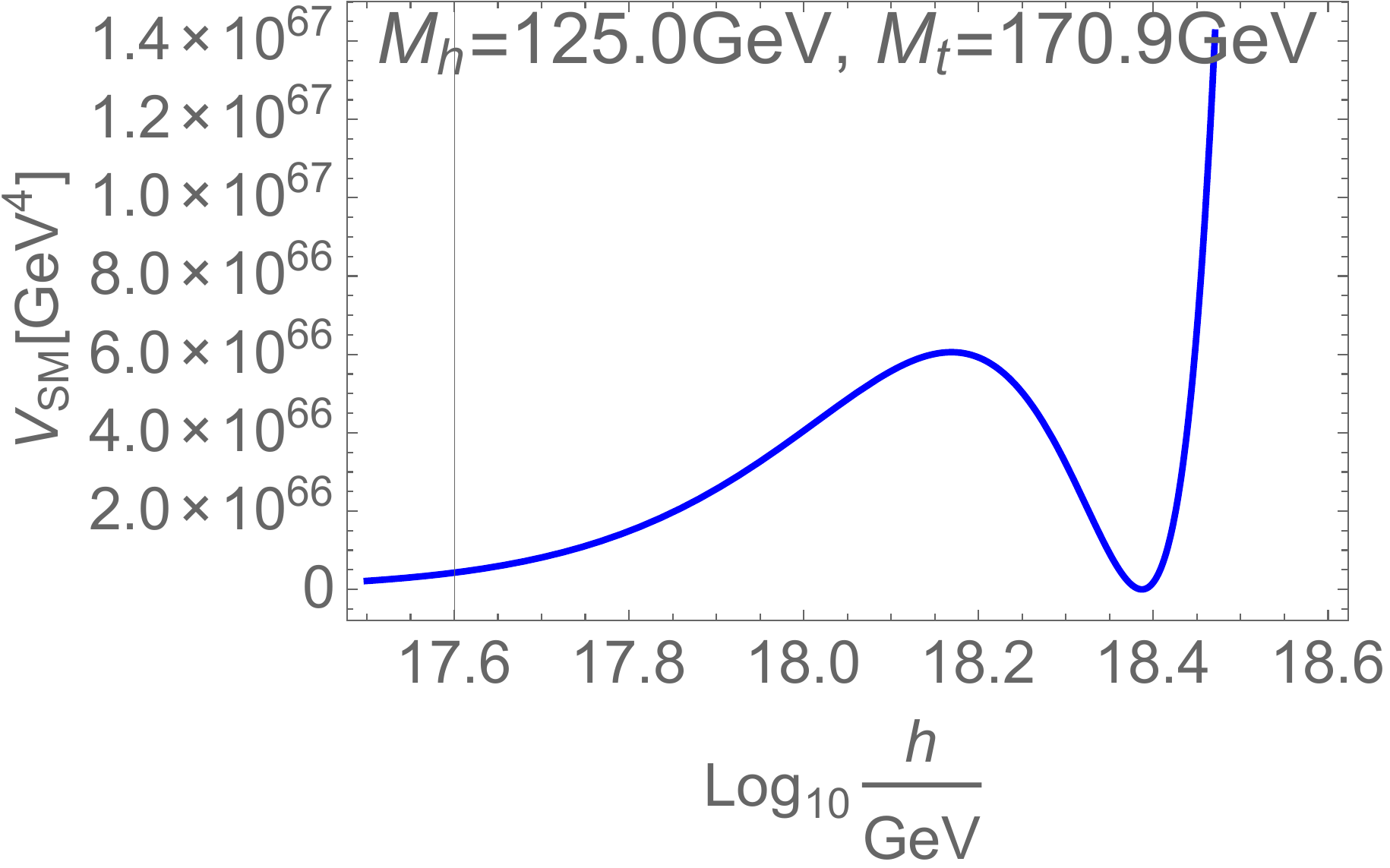}
\end{center}
\caption{The Higgs potential in the SM. Here, the top mass is fine-tuned so that another vacuum appears around the Planck scale.
}
\label{fig:higgs}
\end{figure}

The Coleman's theory \cite{Coleman:1988tj} is one of the promising candidates for solving the naturalness problem, and recently attracts much attention because it is formulated in the Lorentzian framework \cite{MEP,Hamada:2015dja}. It predicts that the parameters (couplings) $\overrightarrow{\lambda}$ of local field theory are fixed at the point that strongly dominates in the integration of the partition function of the universe:
\be Z
\sim\int d\overrightarrow{\lambda} f(\overrightarrow{\lambda})\exp\left(-i\varepsilon(\overrightarrow{\lambda})V_4^{}\right),
\label{eq:partition}
\e
where $f(\overrightarrow{\lambda})$ is an ordinary function of $\overrightarrow{\lambda}$, $\varepsilon(\overrightarrow{\lambda})$ is the vacuum energy density of a system, and $V_4^{}$ is the spacetime volume. From this equation, one can easily understand that a special point such as a saddle point 
of $\varepsilon(\overrightarrow{\lambda})$ strongly dominates because $V_4^{}$ is quite large. See \cite{MEP,Hamada:2015dja} for more details.

In particular, in the recent work \cite{Hamada:2015dja}, it was shown that the Multiple point criticality principle (MPP) \cite{MPP} can be derived by using the Coleman's theory. The principle is a hypothesis 
which claims that if there are two vacua in a model, the parameters of the model are fixed so that they become degenerate. Of course, this principle originates from the SM Higgs potential.  See Fig.\ref{fig:higgs} for example. This is the Higgs potential of the SM by using the observed Higgs mass $\sim 125$ GeV, and one can see that there is another degenerate vacuum around the Planck scale.

%

In the following, we show that the CC in the SM can be also derived by using Eq.(\ref{eq:partition}). We first review the result in \cite{Hamada:2015dja} because the MPP plays an important role for the following discussion.
\\

\begin{figure}
\begin{center}
\includegraphics[width=8.5cm]{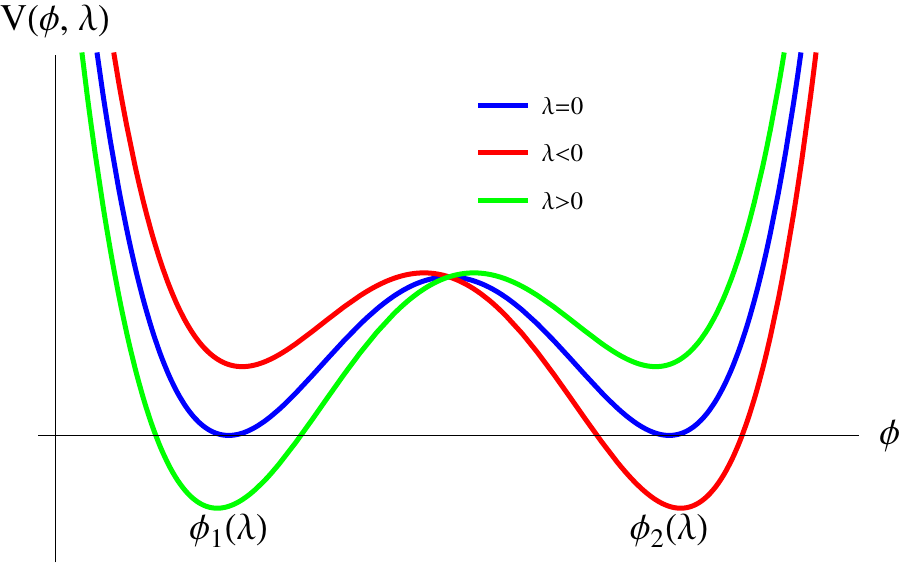}
\end{center}
\caption{Schematic behavior of $V(\phi,\lambda)$. The blue line corresponds to $\lambda=0$, and the green (red) line corresponds to $\lambda>0$ $(<0)$.}
\label{fig:MPP}
\end{figure}

Let us first consider a general potential $V(\phi,\lambda)$ of a scalar field $\phi$ having two minima at $\phi_{1}(\lambda)$ and $\phi_{2}(\lambda)$, where we take $\phi_{1}(\lambda)<\phi_{2}(\lambda)$. Here, $\lambda$ is one of the coupling constants of the theory. For example, it corresponds to the Higgs quartic coupling or the top Yukawa in the SM. We assume that two minima become degenerate when $\lambda$ is equal to zero
, and that the signature of $\lambda$ is chosen as 
\aln{V(\phi_{1}(\lambda),\lambda)&<V(\phi_{2}(\lambda),\lambda)\nonumber\text{ \h{5mm}for $\lambda>0$,}\\
\\
V(\phi_{1}(\lambda),\lambda)&>V(\phi_{2}(\lambda),\lambda)\nonumber\text{ \h{5mm}for $\lambda<0$}.}
See Fig.\ref{fig:MPP} for example. Then, the true vacuum expectation value $\phi_{\text{vac}}(\lambda)$ and the vacuum energy density $\varepsilon(\lambda)$ are given by 
\be \phi_{\text{vac}}(\lambda)=\begin{cases}\phi_{2}(\lambda)&\text{ for $\lambda<0$}\\
\phi_{1}(\lambda)&\text{ for $\lambda>0$}\end{cases}\h{3mm},\h{3mm}
\varepsilon(\lambda)=\begin{cases}V(\phi_{2}(\lambda))&\text{for $\lambda<0$}\\
V(\phi_{1}(\lambda))&\text{for $\lambda>0$}.\end{cases}\label{eq:vac2}\e
Now, in order to examine the dominant point in Eq.(\ref{eq:partition}), let us use the following mathematical formula: If $g(\lambda)$ is smooth and monotonic in the $\lambda>0$ region, and $g'(0)\neq0$, we have 
\be e^{ikg(\lambda)}\theta(\lambda)\underset{k\rightarrow\infty}{\sim}\frac{i}{k}\left(\frac{dg}{d\lambda}\right)^{-1}e^{ikg(0)}\delta(\lambda),\label{eq:delta}\e
where $\theta(\lambda)$ is a step function. 
The proof is as follows. By multiplying a test function $F(\lambda)$ with finite support to $e^{ikg(\lambda)}$, and integrating from 0 to $\infty$, we obtain
\aln{ \int_{0}^{\infty}d\lambda e^{ig(\lambda)k}F(\lambda)&=\int_{g(0)}^{\infty}dg\h{1mm}\left(\frac{dg}{d\lambda}\right)^{-1}e^{ikg}F(\lambda=\lambda(g))\nonumber\\
&=\left[\frac{e^{ikg}}{ik}\left(\frac{dg}{d\lambda}\right)^{-1}F(\lambda(g))\right]_{g(0)}^{\infty}+{\cal{O}}\left(\frac{1}{k^{2}}\right)\nonumber\\
&=\frac{i}{k}\left(\frac{dg}{d\lambda}\right)^{-1} e^{ikg(0)}F(\lambda)\Biggl|_{\lambda=0}+{\cal{O}}\left(\frac{1}{k^{2}}\right).\label{eq:delta1}}
Thus, Eq.(\ref{eq:delta}) holds in the $k\rightarrow\infty$ limit. 

\begin{figure}
\begin{center}
\begin{tabular}{c}
\begin{minipage}{0.5\hsize}
\begin{center}
\includegraphics[width=7.5cm]{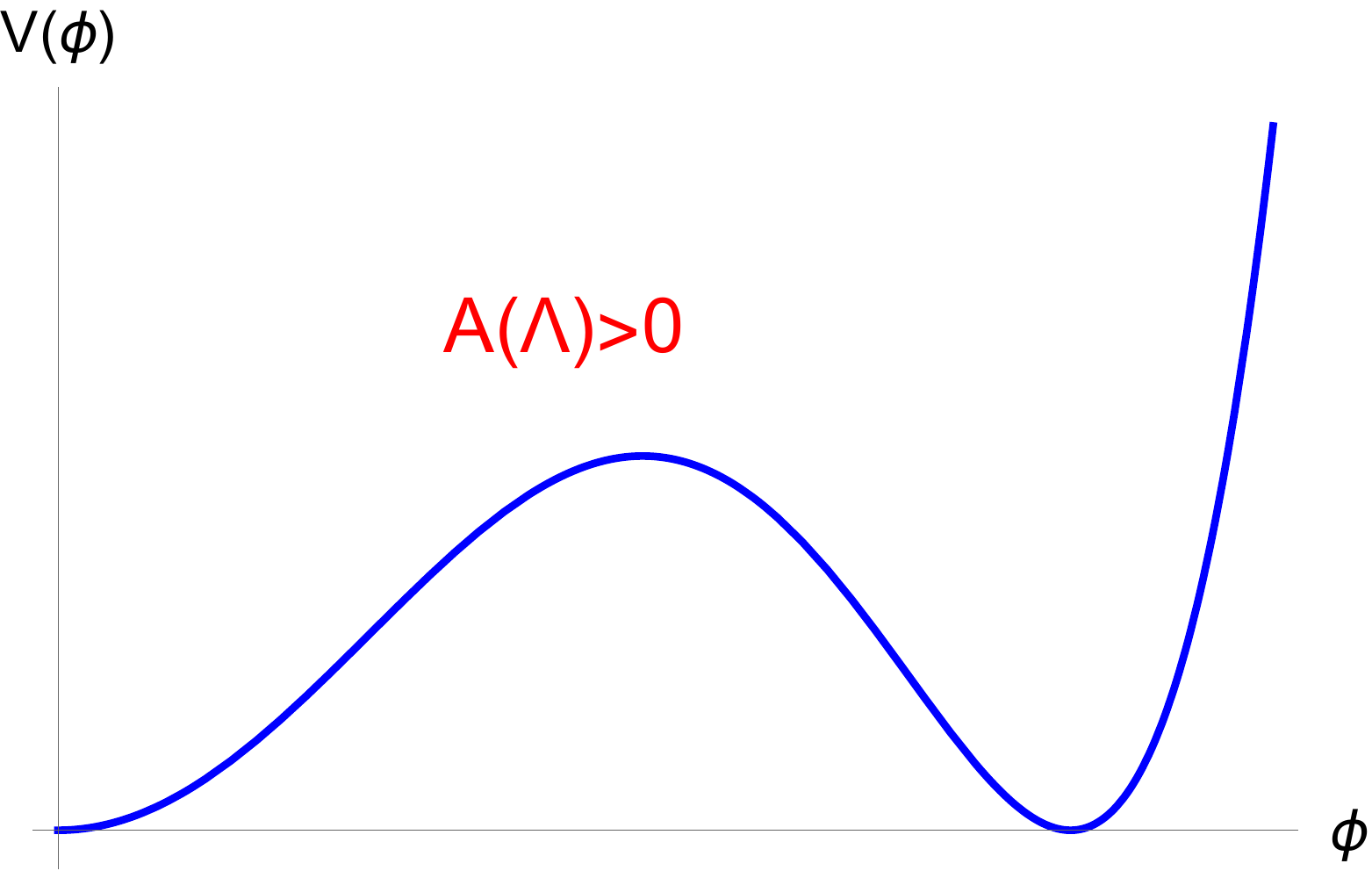}
\end{center}
\end{minipage}
\begin{minipage}{0.5\hsize}
\begin{center}
\includegraphics[width=7.5cm]{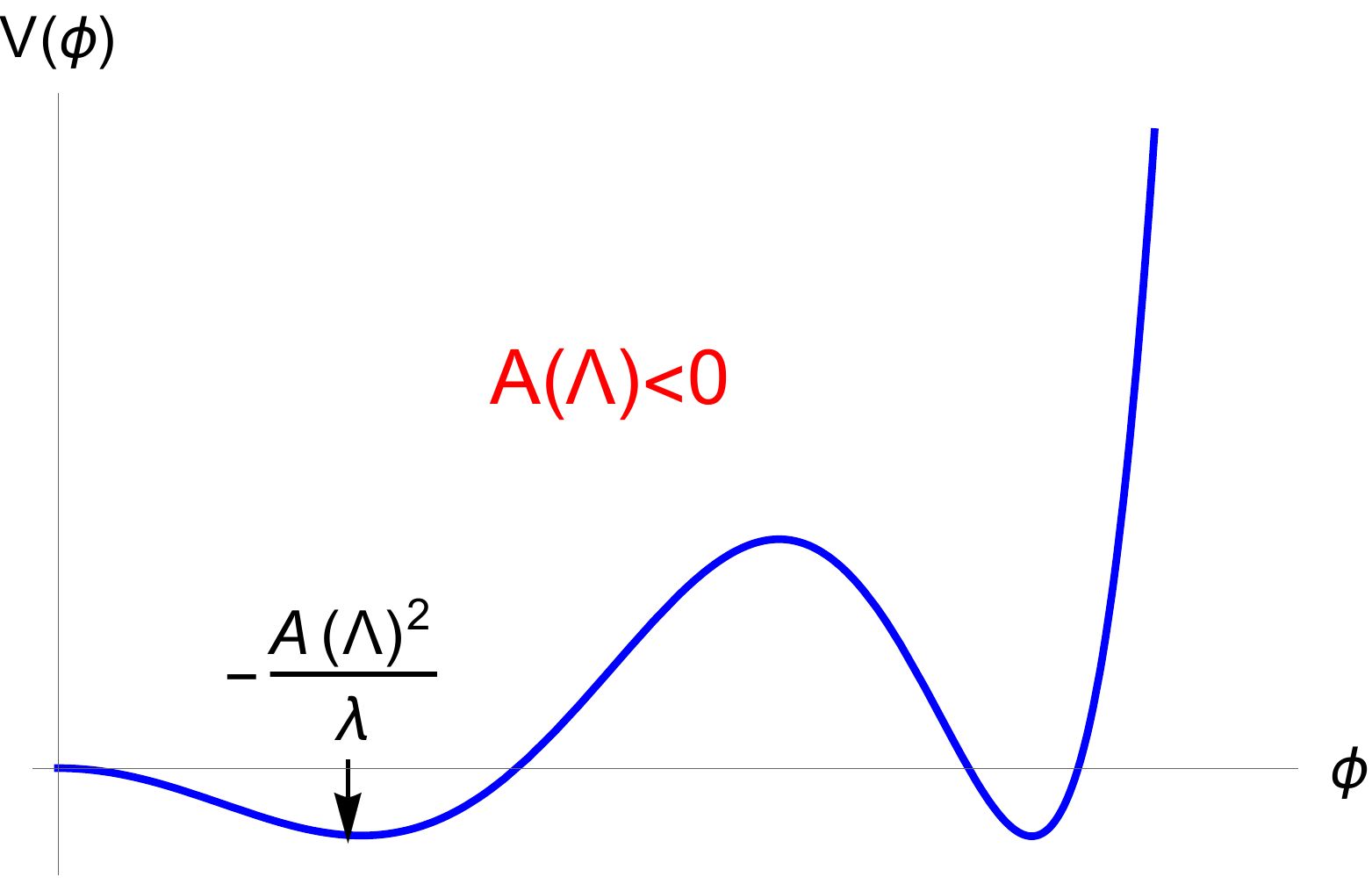}
\end{center}
\end{minipage}
\end{tabular}
\end{center}
\caption{The typical behaviors of the Higgs potential where the MPP is realized. Left (Right) panel shows the $A(\Lambda)>0$ $(<0)$ case.
}
\label{fig:potential}
\end{figure}

By assuming that $V(\phi_{1}(\lambda))$ and $V(\phi_{2}(\lambda))$ are monotonic in $\lambda>0$ and $\lambda<0$ respectively, and using Eq.(\ref{eq:delta}), we have
\be 
e^{-i\varepsilon(\lambda)V_{4}}\sim-\frac{ie^{-i\varepsilon(0)V_{4}}}{V_{4}}\times\left[\left(\frac{V(\phi_{1}(\lambda))}{d\lambda}\right)^{-1}\Biggl|_{0+}-\left(\frac{V(\phi_{2}(\lambda))}{d\lambda}\right)^{-1}\Biggl|_{0-}\right]\delta(\lambda),
\e
which leads to 
\be 
Z\sim\frac{f(0)}{V_{4}}\times e^{-i\varepsilon(0)V_{4}}.
\e
This is the derivation of the MPP in the context of the Coleman's theory. In the SM case, the Higgs potential can have actually two degenerate vacua with vanishing vacuum energy. (See again Fig.\ref{fig:higgs}.)
\\

Let us now derive the CC by assuming that the MPP is always realized from the above argument. In the SM, because $A(\mu)$ is proportional to $A(\Lambda)$, we treat the latter as a free parameter in the following. In this case, the vacuum energy density is given by
\be
\varepsilon(A(\Lambda))=\begin{cases}0&\text{for $A(\Lambda)>0$}\\
-\frac{A(\Lambda)^2}{\lambda}&\text{for $A(\Lambda)<0$}\end{cases},
\label{eq:vac2}
\e
where $\lambda$ is the effective Higgs quartic coupling at the weak scale $\sim{\cal{O}}(0.1)$. See Fig.\ref{fig:potential} for example. Therefore, by using Eq.(\ref{eq:delta}), we can see that $A(\Lambda)=0$ dominates at least in the $A(\Lambda)<0$ region, wheres there seems no dominant point in the $A(\Lambda)>0$ region. However, in addition to the above vacuum energy, 
there is always a contribution from the zero point oscillation energy:
\be \varepsilon_{\text{zero}}^{}(A(\Lambda))=\int\frac{d^3\mathbf{k}}{(2\pi)^3} \sqrt{\mathbf{k}^2+c A(\Lambda)},
\e 
where $c$ is a positive ${\cal{O}}(1)$ coefficient determined by the RGE of the SM. This is apparently a monotonic function of $\Lambda(\Lambda)$, so we can conclude that $A(\Lambda)=0$ also dominates even in the $A(\Lambda)>0$ region
\footnote{The zero point energy also exists in the $A(\Lambda)<0$ region. However, this does not change our conclusion because it is also monotonic.}. This is the derivation of the CC in the SM. Although we have not considered the extensions of the SM here, it might be interesting to study whether the CC is also realizable even in such extensions while preserving their phenomenological aspects such as the EWSB.

\section*{Acknowledgement} 
This work is supported by the Grant-in-Aid for Japan Society for the Promotion of Science (JSPS) Fellows No.27$\cdot$1771.

\appendix 
\def\thesection{Appendix \Alph{section}}

\end{document}